\documentclass[10pt]{elsarticle}
\usepackage{lineno,hyperref}
\usepackage{amsmath,amssymb}
\usepackage{hyperref}
\hypersetup{
    colorlinks=true,
    linkcolor=blue,
}
\usepackage[figurename=Fig., tablename=Table]{caption}
\usepackage{color}
\usepackage{cases}
\usepackage{comment}
\usepackage{ulem}
\usepackage{stfloats}
\usepackage[top=30truemm,bottom=30truemm,left=15truemm,right=15truemm]{geometry}
\journal{Journal of \LaTeX\ Templates}
\begin{document}
\begin{frontmatter}

\title{Search for solar electron anti-neutrinos due to spin-flavor precession in the Sun with Super-Kamiokande-IV
}

\newcommand{\AFFicrr}{\address[AFFicrr]{Kamioka Observatory, Institute for Cosmic Ray Research, University of Tokyo, Kamioka, Gifu 506-1205, Japan}}
\newcommand{\AFFkashiwa}{\address[AFFkashiwa]{Research Center for Cosmic Neutrinos, Institute for Cosmic Ray Research, University of Tokyo, Kashiwa, Chiba 277-8582, Japan}}
\newcommand{\AFFipmu}{\address[AFFipmu]{Kavli Institute for the Physics and Mathematics of the Universe (WPI), The University of Tokyo Institutes for Advanced Study, University of Tokyo, Kashiwa, Chiba 277-8583, Japan }}
\newcommand{\AFFmad}{\address[AFFmad]{Department of Theoretical Physics, University Autonoma Madrid, 28049 Madrid, Spain}}
\newcommand{\AFFbu}{\address[AFFbu]{Department of Physics, Boston University, Boston, MA 02215, USA}}
\newcommand{\AFFuci}{\address[AFFuci]{Department of Physics and Astronomy, University of California, Irvine, Irvine, CA 92697-4575, USA }}
\newcommand{\AFFcsu}{\address[AFFcsu]{Department of Physics, California State University, Dominguez Hills, Carson, CA 90747, USA}}
\newcommand{\AFFcnm}{\address[AFFcnm]{Institute for Universe and Elementary Particles, Chonnam National University, Gwangju 61186, Korea}}
\newcommand{\AFFduke}{\address[AFFduke]{Department of Physics, Duke University, Durham NC 27708, USA}}
\newcommand{\AFFfukuoka}{\address[AFFfukuoka]{Junior College, Fukuoka Institute of Technology, Fukuoka, Fukuoka 811-0295, Japan}}
\newcommand{\AFFgifu}{\address[AFFgifu]{Department of Physics, Gifu University, Gifu, Gifu 501-1193, Japan}}
\newcommand{\AFFgist}{\address[AFFgist]{GIST College, Gwangju Institute of Science and Technology, Gwangju 500-712, Korea}}
\newcommand{\AFFuh}{\address[AFFuh]{Department of Physics and Astronomy, University of Hawaii, Honolulu, HI 96822, USA}}
\newcommand{\AFFicl}{\address[AFFicl]{Department of Physics, Imperial College London , London, SW7 2AZ, United Kingdom }}
\newcommand{\AFFkek}{\address[AFFkek]{High Energy Accelerator Research Organization (KEK), Tsukuba, Ibaraki 305-0801, Japan }}
\newcommand{\AFFkobe}{\address[AFFkobe]{Department of Physics, Kobe University, Kobe, Hyogo 657-8501, Japan}}
\newcommand{\AFFkyoto}{\address[AFFkyoto]{Department of Physics, Kyoto University, Kyoto, Kyoto 606-8502, Japan}}
\newcommand{\AFFliv}{\address[AFFliv]{Department of Physics, University of Liverpool, Liverpool, L69 7ZE, United Kingdom}}
\newcommand{\AFFmiyagi}{\address[AFFmiyagi]{Department of Physics, Miyagi University of Education, Sendai, Miyagi 980-0845, Japan}}
\newcommand{\AFFnagoya}{\address[AFFnagoya]{Institute for Space-Earth Environmental Research, Nagoya University, Nagoya, Aichi 464-8602, Japan}}
\newcommand{\AFFkmi}{\address[AFFkmi]{Kobayashi-Maskawa Institute for the Origin of Particles and the Universe, Nagoya University, Nagoya, Aichi 464-8602, Japan}}
\newcommand{\AFFpol}{\address[AFFpol]{National Centre For Nuclear Research, 02-093 Warsaw, Poland}}
\newcommand{\AFFsuny}{\address[AFFsuny]{Department of Physics and Astronomy, State University of New York at Stony Brook, NY 11794-3800, USA}}
\newcommand{\AFFokayama}{\address[AFFokayama]{Department of Physics, Okayama University, Okayama, Okayama 700-8530, Japan }}
\newcommand{\AFFosaka}{\address[AFFosaka]{Department of Physics, Osaka University, Toyonaka, Osaka 560-0043, Japan}}
\newcommand{\AFFox}{\address[AFFox]{Department of Physics, Oxford University, Oxford, OX1 3PU, United Kingdom}}
\newcommand{\AFFseoul}{\address[AFFseoul]{Department of Physics, Seoul National University, Seoul 151-742, Korea}}
\newcommand{\AFFsheff}{\address[AFFsheff]{Department of Physics and Astronomy, University of Sheffield, S3 7RH, Sheffield, United Kingdom}}
\newcommand{\AFFshizuokasc}{\address[AFFshizuokasc]{Department of Informatics in Social Welfare, Shizuoka University of Welfare, Yaizu, Shizuoka, 425-8611, Japan}}
\newcommand{\AFFstfc}{\address[AFFstfc]{STFC, Rutherford Appleton Laboratory, Harwell Oxford, and Daresbury Laboratory, Warrington, OX11 0QX, United Kingdom}}
\newcommand{\AFFskk}{\address[AFFskk]{Department of Physics, Sungkyunkwan University, Suwon 440-746, Korea}}
\newcommand{\AFFtokyo}{\address[AFFtokyo]{The University of Tokyo, Bunkyo, Tokyo 113-0033, Japan }}
\newcommand{\AFFtodai}{\address[AFFtodai]{Department of Physics, University of Tokyo, Bunkyo, Tokyo 113-0033, Japan }}
\newcommand{\AFFtit}{\address[AFFtit]{Department of Physics,Tokyo Institute of Technology, Meguro, Tokyo 152-8551, Japan }}
\newcommand{\AFFtus}{\address[AFFtus]{Department of Physics, Faculty of Science and Technology, Tokyo University of Science, Noda, Chiba 278-8510, Japan }}
\newcommand{\AFFtoronto}{\address[AFFtoronto]{Department of Physics, University of Toronto, ON, M5S 1A7, Canada }}
\newcommand{\AFFtriumf}{\address[AFFtriumf]{TRIUMF, 4004 Wesbrook Mall, Vancouver, BC, V6T2A3, Canada }}
\newcommand{\AFFtokai}{\address[AFFtokai]{Department of Physics, Tokai University, Hiratsuka, Kanagawa 259-1292, Japan}}
\newcommand{\AFFtsinghua}{\address[AFFtsinghua]{Department of Engineering Physics, Tsinghua University, Beijing, 100084, China}}
\newcommand{\AFFynu}{\address[AFFynu]{Department of Physics, Yokohama National University, Yokohama, Kanagawa, 240-8501, Japan}}
\newcommand{\AFFllr}{\address[AFFllr]{Ecole Polytechnique, IN2P3-CNRS, Laboratoire Leprince-Ringuet, F-91120 Palaiseau, France }}
\newcommand{\AFFbari}{\address[AFFbari]{ Dipartimento Interuniversitario di Fisica, INFN Sezione di Bari and Universit\`a e Politecnico di Bari, I-70125, Bari, Italy}}
\newcommand{\AFFnapoli}{\address[AFFnapoli]{Dipartimento di Fisica, INFN Sezione di Napoli and Universit\`a di Napoli, I-80126, Napoli, Italy}}
\newcommand{\AFFroma}{\address[AFFroma]{INFN Sezione di Roma and Universit\`a di Roma ``La Sapienza'', I-00185, Roma, Italy}}
\newcommand{\AFFpadova}{\address[AFFpadova]{Dipartimento di Fisica, INFN Sezione di Padova and Universit\`a di Padova, I-35131, Padova, Italy}}
\newcommand{\AFFkeio}{\address[AFFkeio]{Department of Physics, Keio University, Yokohama, Kanagawa, 223-8522, Japan}}
\newcommand{\AFFwinnipeg}{\address[AFFwinnipeg]{Department of Physics, University of Winnipeg, MB R3J 3L8, Canada }}
\newcommand{\AFFkcl}{\address[AFFkcl]{Department of Physics, King's College London, London, WC2R 2LS, UK }}
\newcommand{\AFFwarwick}{\address[AFFwarwick]{Department of Physics, University of Warwick, Coventry, CV4 7AL, UK }}
\newcommand{\AFFral}{\address[AFFral]{Rutherford Appleton Laboratory, Harwell, Oxford, OX11 0QX, UK }}
\newcommand{\AFFwu}{\address[AFFwu]{Faculty of Physics, University of Warsaw, Warsaw, 02-093, Poland}}
\newcommand{\AFFbcit}{\address[AFFbcit]{Department of Physics, British Columbia Institute of Technology, Burnaby, BC, V5G 3H2, Canada }}

\author[AFFicrr,AFFipmu]{K.~Abe}
\author[AFFicrr]{C.~Bronner}
\author[AFFicrr,AFFipmu]{Y.~Hayato}
\author[AFFicrr]{M.~Ikeda}
\author[AFFicrr]{S.~Imaizumi}
\author[AFFicrr]{H.~Ito
\corref{CorrespondingAuthour}}
\cortext[CorrespondingAuthour]{Corresponding author}
\ead{itoh@km.icrr.u-tokyo.ac.jp}
\author[AFFicrr,AFFipmu]{J.~Kameda}
\author[AFFicrr]{Y.~Kataoka}
\author[AFFicrr,AFFipmu]{M.~Miura} 
\author[AFFicrr,AFFipmu]{S.~Moriyama}
\author[AFFicrr]{Y.~Nagao} 
\author[AFFicrr,AFFipmu]{M.~Nakahata}
\author[AFFicrr,AFFipmu]{Y.~Nakajima}
\author[AFFicrr,AFFipmu]{S.~Nakayama}
\author[AFFicrr]{T.~Okada}
\author[AFFicrr]{K.~Okamoto}
\author[AFFicrr]{A.~Orii}
\author[AFFicrr]{G.~Pronost}
\author[AFFicrr]{H.~Sekiya} 
\author[AFFicrr]{M.~Shiozawa}
\author[AFFicrr]{Y.~Sonoda}
\author[AFFicrr]{Y.~Suzuki} 
\author[AFFicrr,AFFipmu]{A.~Takeda}
\author[AFFicrr]{Y.~Takemoto}
\author[AFFicrr]{A.~Takenaka}
\author[AFFicrr]{H.~Tanaka}
\author[AFFicrr]{T.~Yano}
\author[AFFkashiwa]{R.~Akutsu}
\author[AFFkashiwa]{S.~Han} 
\author[AFFkashiwa,AFFipmu]{T.~Kajita} 
\author[AFFkashiwa,AFFipmu]{K.~Okumura}
\author[AFFkashiwa]{T.~Tashiro}
\author[AFFkashiwa]{R.~Wang}
\author[AFFkashiwa]{J.~Xia}
\author[AFFmad]{D.~Bravo-Bergu\~{n}o}
\author[AFFmad]{L.~Labarga}
\author[AFFmad]{Ll.~Marti}
\author[AFFmad]{B. Zaldivar}
\author[AFFbu]{F.~d.~M.~Blaszczyk}
\author[AFFbu,AFFipmu]{E.~Kearns}
\author[AFFbu]{J.~L.~Raaf}
\author[AFFbu,AFFipmu]{J.~L.~Stone}
\author[AFFbu]{L.~Wan}
\author[AFFbu]{T.~Wester}
\author[AFFbcit]{B.~W.~Pointon}
\author[AFFuci]{J.~Bian}
\author[AFFuci]{N.~J.~Griskevich}
\author[AFFuci]{W.~R.~Kropp}
\author[AFFuci]{S.~Locke} 
\author[AFFuci]{S.~Mine} 
\author[AFFuci,AFFipmu]{M.~B.~Smy}
\author[AFFuci,AFFipmu]{H.~W.~Sobel} 
\author[AFFuci]{V.~Takhistov\fnref{aaa}} \fntext[aaa]{Also at Department of Physics and Astronomy, UCLA, CA 90095-1547, USA.}
\author[AFFuci]{P.~Weatherly} 
\author[AFFcsu]{J.~Hill}
\author[AFFcnm]{J.~Y.~Kim}
\author[AFFcnm]{I.~T.~Lim}
\author[AFFcnm]{R.~G.~Park}
\author[AFFduke]{B.~Bodur}
\author[AFFduke,AFFipmu]{K.~Scholberg}
\author[AFFduke,AFFipmu]{C.~W.~Walter}
\author[AFFllr]{L.~Bernard} 
\author[AFFllr]{A.~Coffani}
\author[AFFllr]{O.~Drapier}
\author[AFFllr]{S.~El Hedri}
\author[AFFllr]{A.~Giampaolo}
\author[AFFllr]{M.~Gonin}
\author[AFFllr]{Th.~A.~Mueller}
\author[AFFllr]{P.~Paganini}
\author[AFFllr]{B.~Quilain}
\author[AFFfukuoka]{T.~Ishizuka}
\author[AFFgifu]{T.~Nakamura}
\author[AFFgist]{J.~S.~Jang}
\author[AFFuh]{J.~G.~Learned} 
\author[AFFicl]{L.~H.~V.~Anthony}
\author[AFFicl]{A.~A.~Sztuc} 
\author[AFFicl]{Y.~Uchida}
\author[AFFbari]{V.~Berardi}
\author[AFFbari]{M.~G.~Catanesi}
\author[AFFbari]{E.~Radicioni}
\author[AFFnapoli]{N.~F.~Calabria}
\author[AFFnapoli]{L.~N.~Machado}
\author[AFFnapoli]{G.~De Rosa}
\author[AFFpadova]{G.~Collazuol}
\author[AFFpadova]{F.~Iacob}
\author[AFFpadova]{M.~Lamoureux}
\author[AFFpadova]{N.~Ospina}
\author[AFFroma]{L.\,Ludovici}
\author[AFFkeio]{Y.~Nishimura}
\author[AFFkek]{S.~Cao}
\author[AFFkek]{M.~Friend}
\author[AFFkek]{T.~Hasegawa} 
\author[AFFkek]{T.~Ishida} 
\author[AFFkek]{M.~Jakkapu} 
\author[AFFkek]{T.~Kobayashi} 
\author[AFFkek]{T.~Matsubara}
\author[AFFkek]{T.~Nakadaira} 
\author[AFFkek,AFFipmu]{K.~Nakamura}
\author[AFFkek]{Y.~Oyama} 
\author[AFFkek]{K.~Sakashita} 
\author[AFFkek]{T.~Sekiguchi} 
\author[AFFkek]{T.~Tsukamoto}
\author[AFFkobe]{Y.~Nakano}
\author[AFFkobe]{T.~Shiozawa}
\author[AFFkobe]{A.~T.~Suzuki}
\author[AFFkobe,AFFipmu]{Y.~Takeuchi}
\author[AFFkobe]{S.~Yamamoto}
\author[AFFkyoto]{A.~Ali}
\author[AFFkyoto]{Y.~Ashida}
\author[AFFkyoto]{J.~Feng}
\author[AFFkyoto]{S.~Hirota}
\author[AFFkyoto]{A.~K.~Ichikawa}
\author[AFFkyoto]{T.~Kikawa}
\author[AFFkyoto]{M.~Mori}
\author[AFFkyoto,AFFipmu]{T.~Nakaya}
\author[AFFkyoto,AFFipmu]{R.~A.~Wendell}
\author[AFFkyoto]{Y.~Yasutome}
\author[AFFliv]{P.~Fernandez}
\author[AFFliv]{N.~McCauley}
\author[AFFliv]{P.~Mehta}
\author[AFFliv]{A.~Pritchard}
\author[AFFliv]{K.~M.~Tsui}
\author[AFFmiyagi]{Y.~Fukuda}
\author[AFFnagoya,AFFkmi]{Y.~Itow}
\author[AFFnagoya]{H.~Menjo}
\author[AFFnagoya]{T.~Niwa}
\author[AFFnagoya]{K.~Sato}
\author[AFFnagoya]{M.~Tsukada}
\author[AFFpol]{P.~Mijakowski}
\author[AFFsuny]{C.~K.~Jung}
\author[AFFsuny]{C.~Vilela}
\author[AFFsuny]{M.~J.~Wilking}
\author[AFFsuny]{C.~Yanagisawa\fnref{ccc}} \fntext[ccc]{Also at BMCC/CUNY, Science Department, New York, New York, 10007, USA.}
\author[AFFokayama]{M.~Harada}
\author[AFFokayama]{K.~Hagiwara}
\author[AFFokayama]{T.~Horai}
\author[AFFokayama]{H.~Ishino}
\author[AFFokayama]{S.~Ito}
\author[AFFokayama,AFFipmu]{Y.~Koshio}
\author[AFFokayama]{W.~Ma}
\author[AFFokayama]{N.~Piplani}
\author[AFFokayama]{S.~Sakai}
\author[AFFosaka]{Y.~Kuno}
\author[AFFox]{G.~Barr}
\author[AFFox]{D.~Barrow}
\author[AFFox,AFFipmu]{L.~Cook}
\author[AFFox]{A.~Goldsack}
\author[AFFox]{S.~Samani}
\author[AFFox,AFFipmu]{C.~Simpson}
\author[AFFox,AFFstfc]{D.~Wark}
\author[AFFral]{F.~Nova}
\author[AFFkcl]{T.~Boschi}
\author[AFFkcl]{F.~Di Lodovico}
\author[AFFkcl]{M.~Taani}
\author[AFFkcl]{J.~Migenda}
\author[AFFkcl]{S.~Molina Sedgwick\fnref{ddd}} \fntext[ddd]{Currently at Queen Mary University of London, London, E1 4NS, United Kingdom.}
\author[AFFkcl]{S.~Zsoldos}
\author[AFFseoul]{J.~Y.~Yang}
\author[AFFsheff]{S.~J.~Jenkins}
\author[AFFsheff]{M.~Malek}
\author[AFFsheff]{J.~M.~McElwee}
\author[AFFsheff]{O.~Stone}
\author[AFFsheff]{M.~D.~Thiesse}
\author[AFFsheff]{L.~F.~Thompson}
\author[AFFshizuokasc]{H.~Okazawa}
\author[AFFskk]{S.~B.~Kim}
\author[AFFskk]{I.~Yu}
\author[AFFtokai]{K.~Nishijima}
\author[AFFtokyo]{M.~Koshiba\fnref{koshiba}} \fntext[koshiba]{Deceased.}
\author[AFFtodai]{K.~Iwamoto}
\author[AFFtodai]{N.~Ogawa}
\author[AFFtodai,AFFipmu]{M.~Yokoyama}
\author[AFFipmu]{K.~Martens}
\author[AFFipmu,AFFuci]{M.~R.~Vagins}
\author[AFFtit]{S.~Izumiyama}
\author[AFFtit]{M.~Kuze}
\author[AFFtit]{M.~Tanaka}
\author[AFFtit]{T.~Yoshida}
\author[AFFtus]{M.~Inomoto}
\author[AFFtus]{M.~Ishitsuka}
\author[AFFtus]{R.~Matsumoto}
\author[AFFtus]{K.~Ohta}
\author[AFFtus]{M.~Shinoki}
\author[AFFtoronto]{J.~F.~Martin}
\author[AFFtoronto]{H.~A.~Tanaka}
\author[AFFtoronto]{T.~Towstego}
\author[AFFtriumf]{M.~Hartz}
\author[AFFtriumf]{A.~Konaka}
\author[AFFtriumf]{P.~de Perio}
\author[AFFtriumf]{N.~W.~Prouse}
\author[AFFtsinghua]{S.~Chen}
\author[AFFtsinghua]{B.~D.~Xu}
\author[AFFwarwick]{B.~Richards}
\author[AFFwinnipeg]{B.~Jamieson}
\author[AFFwinnipeg]{J.~Walker}
\author[AFFynu]{A.~Minamino}
\author[AFFynu]{K.~Okamoto}
\author[AFFynu]{G.~Pintaudi}
\author[AFFynu]{R.~Sasaki}
\author[AFFwu]{and M. Posiadala-Zezula}
\author{(The Super-Kamiokande Collaboration)}

\AFFicrr
\AFFkashiwa
\AFFmad
\AFFbu
\AFFbcit
\AFFuci
\AFFcsu
\AFFcnm
\AFFduke
\AFFllr
\AFFfukuoka
\AFFgifu
\AFFgist
\AFFuh
\AFFicl
\AFFbari
\AFFnapoli
\AFFpadova
\AFFroma
\AFFkcl
\AFFkeio
\AFFkek
\AFFkobe
\AFFkyoto
\AFFliv
\AFFmiyagi
\AFFnagoya
\AFFkmi
\AFFpol
\AFFsuny
\AFFokayama
\AFFosaka
\AFFox
\AFFral
\AFFseoul
\AFFsheff
\AFFshizuokasc
\AFFstfc
\AFFskk
\AFFtokai
\AFFtokyo
\AFFtodai
\AFFipmu
\AFFtit
\AFFtus
\AFFtoronto
\AFFtriumf
\AFFtsinghua
\AFFwarwick
\AFFwinnipeg
\AFFynu
\AFFwu

\begin{abstract}
Due to a very low production rate of electron anti-neutrinos ($\bar{\nu}_e$) via nuclear fusion in the Sun, a flux of solar $\bar{\nu}_e$ is unexpected. An appearance of $\bar{\nu}_e$ in solar neutrino flux opens a new window for the new physics beyond the standard model. In particular, a spin-flavor precession process is expected to convert an electron neutrino into an electron anti-neutrino  (${\nu_e\to\bar{\nu}_e}$) when neutrino has a finite magnetic moment. In this work, we have searched for solar $\bar{\nu}_e$ in the Super-Kamiokande experiment, using neutron tagging to identify their inverse beta decay signature. We identified 78 $\bar{\nu}_e$ candidates for neutrino energies of 9.3 to 17.3~MeV in 2970.1 live days with a fiducial volume of 22.5 kiloton water (183.0~kton$\cdot$year exposure). The energy spectrum has been consistent with background predictions and we thus derived a 90\% confidence level upper limit of ${4.7\times10^{-4}}$ on the $\nu_e\to\bar{\nu}_e$ conversion probability in the Sun. We used this result to evaluate the sensitivity of future experiments, notably the Super-Kamiokande Gadolinium (SK-Gd) upgrade.
\end{abstract}


\end{frontmatter}

\begin{twocolumn}
\section{Introduction}

While the Sun is known to produce neutrinos through nuclear fusion processes abundantly, small amounts of antineutrinos can also be emitted through multiple channels. In 1990, Malaney et al. \cite{Malaney1990} predicted that electron antineutrinos ($\bar{\nu}_e$'s) could be produced in the Sun through the following processes: (1) $\beta^-$ decays of radioactive elements such as $^{40}$K (neutrino energy less than $1.4~\rm MeV$, flux ${\sim200\;{\rm cm^{-2}\;s^{-1}}}$ at Earth's surface), (2) $\beta^-$ decays following the photo-fission of heavy isotopes such as $^{238}$U and $^{232}$Th (neutrino energy of 3--9~MeV, flux $\sim$ $\rm10^{-3}~ cm^{-2} s^{-1}$). To date, none of these antineutrinos have been observed. However, the fact that the fluxes predicted by the Standard Solar Model are minimal makes solar antineutrinos a powerful probe of new physics. In 2009, D$\acute{\i}$az et al. showed that a non-zero second order term of the neutrino-antineutrino conversion probability, $P_{\nu\to\bar{\nu}}$, would be a distinctive Lorentz violation \cite{DAIZ2009}. Furthermore, in 2003, Akhmedov and Pulido calculated the probability of $\nu_{e}^{L}\to{\bar{\nu}}_{e}^{R}$ conversion caused by spin-flavour precession in the Sun (lepton-number nonconservation) and ordinary oscillation processes on the way from the Sun to the Earth \cite{Akhmedov2003},
\begin{eqnarray}
&P_{\nu_{e}\to\bar{\nu}_e} 
&\sim 1.8 \times 10^{-10} \sin^2{2\theta_{12}} \nonumber \\
&&\times \biggl[ \frac{\mu}{10^{-12} \mu_B} \frac{B_T (0.05 R_{\odot} ) } {10 \; {\rm kG} } \biggr]^2,\label{eq.1.1}
\end{eqnarray}
where $\theta_{12}=34.5{^\circ}^{+1.2}_{-1.0}$ \cite{Salas2018} is a component of the neutrino oscillation mixing angles, $\mu_B$ is the Bohr magneton, ${\mu<2.9\times10^{-11}\mu_B}$ \cite{GEMMA2013} is the neutrino magnetic moment, and $B_T (r)$ is the solar magnetic field at $r=0.05 R_{\odot}$. The magnetic field inside the Sun is poorly characterized, and can range from $\sim$600~G \cite{Kitchatinov2008} to $\sim$7~MG \cite{Friedland2003} in the radiation zone of the Sun.

Until now the KamLAND experiment set the tightest constraint on $P_{\nu_e\to\bar{\nu}_e}$ with an upper limit of $5.3\times10^{-5}$ at 90\% confidence level (C.L.) in the 8.3--31.8~MeV neutrino energy range, with 4.53 kton$\cdot$years exposure (2343~live days) \cite{KamLAND2012} assuming an unoscillated ${\rm ^{8}B}$ neutrino flux of $5.88\times10^6~{\rm cm^{-2} s^{-1}}$ \cite{Serenelli2010}. Also, the Borexino experiment reported a solar $\bar{\nu}_e$ flux limit of $384~{\rm cm^{-2} s^{-1}}$ at $90\%~\rm C.L.$ in the neutrino energy region of 1.8--16.8~MeV after 2485 live days, which corresponds to $P_{\nu_e\to\bar{\nu}_e}<7.2\times10^{-5}$ at 90\%~\rm C.L. \cite{Borexino2021}. Both experiments identified $\bar{\nu}_e$ events by tagging both the neutron and the positron from inverse beta decays (IBD), ${\bar{\nu}_e + p \to e^+ + n}$. The IBD events are observed as a sum of scintillation light deposited before the positron stops (i.e. its kinetic energy) and light from two 0.511~MeV annihilation $\gamma$s. The neutron emission is identified by its delayed capture signal, a 2.2~MeV $\gamma$. In these detectors, the mean delay between the prompt event and this capture signal is typically $200~\rm\mu s$, facilitating neutron identification. The main background in these experiments comes from neutral-current interactions of atmospheric neutrinos on carbon nuclei. 

On the other hand, water Cherenkov detectors have different backgrounds from the liquid scintillator detectors, that is atmospheric neutrino interaction on oxygen nuclei. The hydrogen concentration in water is also different from the liquid scintillator. Thus it is important to perform the solar $\bar{\nu}_e$ search by both detectors. The SNO experiment searched for solar $\bar{\nu}_e$ in a heavy water Cherenkov detector, where $\bar{\nu}_e$ can be detected via the charged-current reaction on deuterium, $\bar{\nu}_e+d \to e^+ + n + n$. For this channel, SNO reported an upper limit on the $\bar{\nu}_e$ flux from the Sun of $\phi_{\bar{\nu}_e}<3.4\times10^{4}~{\rm cm^{-2} s^{-1}}$ $(90\%~\rm C.L.)$ in the 4--14.8~MeV energy range after 305.9~live days, which corresponds to $P_{\nu_e\to\bar{\nu}_e}<8.1\times10^{-3}~(90\%~\rm C.L.)$ \cite{SNO2004}. In water Cherenkov detectors, $\bar{\nu}_e$ events can be detected via the IBD interaction. The first phase of Super-Kamiokande (SK-I) found no significant excess for solar $\bar{\nu}_e$ in selecting events whose directions were not aligned with the direction from the Sun ($\cos\theta_{\rm sun}<0.5$)
\footnote{Defined as the angle between the reconstructed direction of signal candidate and the direction pointing from the Sun. IBD's $e^+$ has almost no directionality from the incoming $\bar{\nu}_e$. This cut is to reject events of solar $\nu_e$ in an elastic scattering interaction.}.
It set an upper limit on the conversion probability of $8\times10^{-3}~(\rm 90\%~C.L.)$ in the 8--20~MeV energy range after 1496~live days \cite{SK2003}. In 2008 for the fourth phase of SK~(SK-IV) the data acquisition (DAQ) system was upgraded \cite{SK-NIMA-2009, SK-IEEE-2010} to detect the delayed signal for 2.2~MeV $\gamma$ emission from neutron capture on hydrogen. The upper limit of the conversion probability in the absence of a signal was calculated to be $4.2\times10^{-4} \; (\rm 90\%~C.L.)$ in 13.3--$31.3~\rm MeV$ and 960 live days \cite{SK2015}.

The next SK phase, called Super-Kamiokande Gadolinium (SK-Gd), will improve the detection efficiency of $\bar{\nu}_e$ via IBD interaction by dissolving gadolinium sulfate into the tank water.

In this work, we present an updated search for solar $\bar{\nu}_e$ in the SK-IV. Compared to the previous search \cite{SK2015}, we here use a more extensive SK-IV data set and an improved neutron tagging procedure using machine learning. The event selection condition is optimized to keep the IBD events efficiently while suppressing the background events. Also, several systematic uncertainties are evaluated to determine $P_{\nu_e\to\bar{\nu}_e}$ in SK-IV, and to perform a realistic estimate of the sensitivity of SK-Gd.

The rest of this article proceeds as follows. In Section~2 we briefly describe the SK detector and its performance. In Section~3 we detail the signal and background simulations used to evaluate our analysis's sensitivity. Then, Sections~4 summarizes the different cuts and the neutron tagging procedure (detailed in Appendix A), while Section~5 describes how we estimate the effects of these cuts on the backgrounds.  Finally, we show our results in Section~6 and discuss a sensitivity evaluation of SK-Gd in Section~7, then conclude.
\section{
\label{sec:SK}
Super-Kamiokande
}
 
SK consists of a stainless steel tank (39.3~m diameter, 41.4~m height), filled with 50~kilotons (kton) of ultra-pure water surrounded by photomultiplier tubes (PMTs). The SK detector consists of two concentric cylindrical volumes separated optically, an inner detector (ID) and an outer detector (OD). We use two kinds of PMTs; 11,129 inward-facing 20-inch PMTs are mounted uniformly on the ID surface and 1,885 outward-facing 8-inch PMTs are mounted uniformly on the OD surface. The details of the SK detector are described elsewhere \cite{Fukuda2003NIM, Abe2014NIM}.

SK started data taking in 1996, and since then has undergone six data-taking phases: SK-I, II, III, IV, V, and SK-Gd (that just started). This search uses data from the SK-IV period, collected between October 2008 and May 2018. Phase IV was characterized by new front-end electronics and a new data processing system \cite{SK-NIMA-2009, SK-IEEE-2010}. For a typical event, data within the time window from $-5$ to $+35~\rm \mu s$ around the trigger time is stored. A trigger relevant for this analysis, called SHE trigger, is issued for events as follows: (a) with more than 70 (58 after September 2011) observed ID PMT hits in a 200~ns time window---equivalent to a 9.5~MeV (7.5~MeV) threshold on the recoil positron kinetic energy---and (b) fewer than 22 OD hits to reject cosmic-ray muon events. In addition to the SHE trigger, an after-trigger (AFT) with a length of 500~$\rm \mu s$ (350~$\rm \mu s$ before November 2008) is issued. These two successive triggers allow to detect the prompt positron signal while providing a 535~$\rm \mu s$ (385~$\rm \mu s$ before November 2008) search window for the delayed 2.2~MeV $\gamma$ from neutron capture. Below the SHE trigger threshold the number of background events sharply increases due to radon's presence of a few $\rm mBq/m^{3}$ level \cite{radon2020} in water, and lowering the threshold would lead to data storage issues. This energy threshold is therefore set by considering both background rates and the speed of data transfer. The analysis presented here considers events with kinetic energies of 7.5--15.5~MeV for a livetime of 2970.1~days (during 2008--2018) and a fiducial volume of 22.5~kton.
\section{Simulation}
\label{section.simulation}

The solar $\bar{\nu}_e$ signal and most of the backgrounds need to be modeled using Monte Carlo (MC) simulations. Here, we present the detail of these simulations for antineutrino IBDs---IBD being both the signal and the irreducible reactor neutrino background---and for backgrounds from atmospheric neutrinos and radioactive decays of $^9$Li. Additionally, the IBD simulation was also used to develop the neutron tagging algorithm detailed in section~\ref{sec.event.4th}.

\subsection{ Solar electron antineutrinos }

The $\bar{\nu}_e$ flux from the Sun is modeled by convolving the $^8$B neutrino flux \cite{winter2006, SNO2013} and the oscillation probability $P_{\nu_e\to\bar{\nu}_e}$. The cross section for IBD interactions is understood and can be calculated according to Ref.~\cite{Strumia2003}. An MC code simulates the associated production of a positron and a neutron. After propagating in water, neutrons are usually captured by hydrogen nuclei near their emission point. Then, the resulting emission of a 2.2~MeV $\gamma$, with a characteristic time constant of $\sim 200~\mu s$, is simulated.

\subsection{ Atmospheric neutrinos }

Atmospheric neutrinos are among the dominant backgrounds in this analysis. The flux of atmospheric neutrinos is predicted by the HKKM2011 model \cite{HKKM2011}. The neutrino-nucleus interaction and subsequent state interactions inside the nucleus are simulated using NEUT 5.3.6 \cite{bib:neut}, i.e. the same interaction model of Ref.~\cite{T2K2019} is used in this study. The initial nucleon momentum distribution follows the spectral function model \cite{bib:benhar1994,bib:benhar2005} for the neutral-current quasielastic (NCQE) interaction and the relativistic Fermi gas model \cite{bib:smithmoniz} for the charged-current quasielastic (CCQE) interaction. CC two-particle-two-hole (2p2h) interactions, where two nucleons participate in the interaction via meson exchange currents, are based on the calculation from Nieves et al. \cite{bib:nieves}. NC 2p2h is not simulated in the current analysis. The BBBA05 and dipole forms \cite{bib:bbba05,bib:kueno} are used to parametrize the vector and axial-vector form factors, respectively. Single-pion production is simulated based on Ref.~\cite{bib:reinsehgal} and a deep inelastic scattering simulation is done using the GRV98 parton distribution function \cite{bib:grv98} with Bodek-Yang corrections \cite{bib:bodekyang}. The final state interactions are simulated with a cascade model. The nuclear de-excitation $\gamma$s are simulated based on the spectroscopic factors calculated by Ankowski et al.~\cite{bib:ankowski}. More detailed descriptions can be found in Ref.~\cite{T2K2019, SK2019NCQE}. One difference from the reference above is about the treatment of the {\it others} state, which is a state affected by short-range correlations or a very high energy excited state. This is included in the ground state in the present analysis. The systematic uncertainties of these nuclear effects are evaluated by replacing the Fermi gas model with the spectral function model, as the cross section uncertainty in Section 5.1.

\subsection{Cosmic-ray induced {\rm $^9$Li}}
\label{sec.sim.li9}

Cosmic-ray muon spallation in the SK detector produces large quantities of radioactive isotopes. These isotopes' decays result in overwhelmingly large spallation backgrounds in the lower energy range of this analysis. Most of these isotopes undergo beta decay, sometimes with $\gamma$ emission, without a neutron and will therefore be efficiently rejected using neutron tagging. However, $^9$Li and $^8$He are the dominant isotopes decay, emitting both an electron and a neutron, mimicking the IBD signal. Actually, $^9$Li events are dominant because $^8$He has lower end-point beta energy and shorter life time than those of $^9$Li \cite{PRC81.025807}. The decay process and production for $^9$Li should therefore be modeled separately. The process of interest here is the beta decay of $^9$Li to $^9$Be, followed by the de-excitation of $^9$Be into $^8$Be with the emission of a neutron \cite{EneLevel_A=8910_2004}. The predicted event rate is calculated as
\begin{eqnarray}
\frac{dN}{dt }= {Y_{\rm ^{9}Li}} \cdot {V_{\rm SK}} \cdot {Br} \cdot
\int
f(E_{\beta}) \varepsilon(E_{\beta}) d{E_{\beta}},
\label{eq.sim.li9}
\end{eqnarray}
where ${Y_{\rm ^{9}Li}=0.86\pm0.12 \; {\rm kton}^{-1} {\rm days}^{-1}}$ \cite{SK2016spallation} is the yield of ${\rm ^{9}Li}$ generated by cosmic-ray muon in the SK, ${V_{\rm SK}=22.5\;{\rm kton}}$ is the fiducial volume, $Br = 0.508 \pm 0.009 $ \cite{EneLevel_A=8910_2004} is the branching ratio of this decay, $f(E_\beta)$ is the simulated energy spectrum as a function of reconstructed kinetic energy $E_{\beta}$,
and $\varepsilon(E_{\beta})$ is the detection efficiency including event selection.

\subsection{ Detector simulation }

A simulation based on GEANT3 \cite{GEANT3} provides detector responses in good agreement with data, which is used to model particle propagation in the water, and the optical properties, photosensor and electronics response in SK. Neutron capture events are weak signals, similar in magnitude to the PMT dark noise. Accurate estimates of this dark noise and its evolution as a function of time are crucial to developing an efficient neutron tagging algorithm. To this end, we use data taken with random trigger timing, utilizing the timing signal for the T2K
\footnote{Tokai-to-Kamioka experiment (T2K) synchronizes timing of neutrino beam injection at Tokai and SK that is 295 km away, using GPS \cite{T2KGPS2011}.}
beam during its beam off periods, so-called T2K dummy data. We then inject this data into simulation results from $18~\rm\mu s$ up to $535~\rm\mu s$ after the positron emission. This injection allows us to account for the effect of dark noise in both the $35~\rm\mu s$ SHE and the $500~\rm\mu s$ AFT triggers. 
\section{Event selection}

In order to select signal-like candidates, data reduction is performed in four steps. Since both this study and the previous supernova relic neutrino (SRN) search \cite{SK2015, SK2012} look for electron antineutrinos, the event selection cuts were applied in a similar way as in the previous with some updated criteria to take into account the specificities of this analysis. The first reduction rejects calibration data and most radioactive background events and was applied with the same cut conditions as in the previous study. The second reduction suppresses muon spallation events. The procedure is same as in the previous study with updated cut criteria. The third reduction is optimized to reduce mainly atmospheric neutrino events. The fourth reduction is the neutron tagging to select IBD candidates and discard accidental coincidences. 

\subsection{Event reconstruction}

In this work, the reconstruction methods used for the vertex ($x$, $y$, and $z$), direction, and energy are the same of Ref.\cite{sk4solar}. The coordinate origin of the vertex is defined as the center of the tank, and we defined the reconstructed radius ($r$), as the cylindrical radius.

\subsection{First reduction}
The first reduction removes bad events and performs noise reduction, where the bad events and noise originate from PMT dark noise, flasher PMTs, and cosmic-ray muons. It was applied with the same cut conditions as in the previous study. It also includes a fiducial volume cut, which corresponds to a fiducial mass of 22.5~kton.

\subsection{Spallation cut}
\label{sec.event.spallcut}
The procedure is same as in the previous study with updated cut criteria.

Cosmic-ray muons produce several short-lived isotopes through interactions with nuclei in the SK water~\cite{Beacom2014, Beacom2015}. These isotopes usually emit electrons or $\gamma$s within the search region (kinetic energy less than 20~MeV) after the muon signal, allowing to eliminate the events effectively.

The second reduction rejects dominant cosmic-ray muon spallation's background events by considering the relation between muon-track and prompt-electron-signal information. In order to confirm a profile of the spallation's events, we have investigated the correlation between a selected event and muons passing through the detector within $\pm 30~\rm s$ of this event. Hereafter, the combined sample of muon-tracks and $e$-signals with the region from $-$30 to 0~s and from 0 to $+$30~s are termed the pre- and post-sample, respectively. 

As outlined in \cite{SK2012}, the time and distance correlations between low energy events and muon tracks can be estimated for spallation events by subtracting pre- and post-sample distributions. Probability density functions (PDFs) are formed for the following variables: the number of muon tracks, maximum $dE/dx$ of a muon track, total deposited charge of a muon track, the distance between the vertex and the muon track, and projected distance along the muon track between the vertex and the point with maximum $dE/dx$. Additionally, to build a random sample that can identify the specific features of the  spallation muons themselves, an electron signal in post-sample and a muon-track signal in toy-MC sample which produced by PDFs are combined.

We then estimate the signal efficiency for these spallation cuts by evaluating the number of random samples before and after cuts. We use the $\cos\theta_{\rm sun}$ distribution, which is the angle between the direction pointing from the Sun and the signal candidate's reconstructed direction. Assuming that the spallation background is flat in $\cos \theta_{\rm sun}$ and that solar $\nu_e$ elastic scattering events are always forward, the number of spallation events can be extracted.

The cut criteria are determined by comparing the likelihood distribution of pre- and post-samples. The MC sample has no contribution from cosmic-ray muons; thus we use the spallation cut's efficiency to evaluate the signal and background events in the MC. The rate of spallation events depends on the electron kinetic energy, and we estimate the signal efficiency ($\varepsilon_{\rm sig,\mu}$) in three kinetic energy regions, 7.5--9.5, 9.5--11.5, and 11.5--19.5~MeV, as a ratio of events before and after the cut procedure for the random sample. We estimate the spallation efficiency using the pre-sample, where efficiency is calculated as a reduction ratio for the spallation events. The spallation ${\rm ^{9}Li}$ events are simulated using a dedicated MC. In order to predict the number of ${\rm ^{9}Li}$ events in the final sample, we apply the spallation cut to the MC sample. The $\rm ^{9}Li$ event efficiency ($\varepsilon_{\rm Li9}$) is derived from $\varepsilon_{\rm sig,\mu}$ and the spallation efficiency. The resulting signal and $^9$Li event efficiencies are summarized in Table~\ref{tab:spacuteff}.

\begin{table}[htb]
\centering
\caption{Signal and $^9$Li event efficiencies of the present spallation cut for each kinetic energy region. The uncertainties come mainly from the statistics of the pre- and post-samples.}
\label{tab:spacuteff}
\begin{tabular}{l c c} \hline 
Kinetic energy region & $\varepsilon_{\rm sig,\mu}$ & $\varepsilon_{\rm Li9}$ \\ \hline
7.5$-$9.5~MeV    & 53.6$\pm$1.6\% & 7.7$\pm$0.2\% \\
9.5$-$11.5~MeV   & 55.2$\pm$1.6\% & 7.6$\pm$0.2\% \\
11.5$-$19.5~MeV  & 75.3$\pm$1.0\% & 16.2$\pm$0.2\% \\ \hline
\end{tabular}
\end{table}

\subsection{Third Reduction}

The third reduction removes atmospheric neutrino backgrounds and remaining radioactive decays using the following criteria.

To further remove backgrounds from the wall, events with ${\it effwall} < \rm 500~cm$ are discarded, where {\it effwall} is the distance from the reconstructed vertex to the ID wall as measured backward along the reconstructed track direction.

Electron events tend to be reconstructed at the Cherenkov angle $\theta_{\rm C} \sim 42^{\circ}$. In contrast, NCQE-like events are often associated with larger angles due to multiple $\gamma$ rings being mis-reconstructed as a single ring by the algorithm. Events are required to satisfy ${38^{\circ} < \theta_{\rm C} < 50^{\circ}}$. The $\bar{\nu}_e$ signal is kept with more than 80--90\% efficiency, while $\sim85\%$ of NC backgrounds are suppressed.

The fuzziness of the Cherenkov ring is characterized by the {\it pilike} parameter, which is defined as follows using an opening angle distribution of all three-hit combinations (triplets) in the event:
\begin{eqnarray}
\label{eq:pilike}
pilike = \frac{N({\rm peak} \pm 3^{\circ})}{N({\rm peak} \pm 10^{\circ})}, 
\end{eqnarray}
\vspace{2truept}
where $N({\rm peak} \pm 3^{\circ})$ and $N({\rm peak} \pm 10^{\circ})$ are the numbers of triplets whose opening angle is within $\pm 3$ and $\pm 10$~degrees of the peak value, respectively. For the solar $\bar{\nu}_e$ signal, $pilike$ peaks around 0.3, while for charged pions and $\gamma$s in NC events it can reach values up to 0.7. The cut removing events with $pilike>0.36$ has kept a signal efficiency of $\sim99\%$.

The total charge detected in a 50~ns time window around the prompt signal, $Q_{50}$, and the number of PMT hits in that time window, $N_{50}$, are calculated. The ratio $Q_{50}/N_{50}$ implies the observed number of photoelectrons per one PMT, so $Q_{50}/N_{50}$ distributions for pion, muon, and electron (or positron) are different. The ratio focuses around 1~p.e. for signal events, while the ratio for pions and muons in atmospheric neutrino events can reach values up to 10~p.e. The cut removing events with $Q_{50}/N_{50}>2~{\rm p.e.}$ has a signal efficiency above 99\%.

\subsection{Neutron Tagging selection}
\label{sec.event.4th}

Although atmospheric neutrinos and spallation backgrounds largely dominate over the signal in the SHE data, these backgrounds can be strongly suppressed by introducing a neutron tagging algorithm to identify IBD events. In the fourth reduction, this algorithm is applied to the surviving events.

The neutron tagging algorithm was developed for an IBD event search in SK-IV \cite{SK2015}. We used the variables calculated from delayed signal hit pattern such as $N_{10}$, $N_{\rm cluster}$, $N_{\rm back}$, and $N_{\rm low}$, referred to Table~\ref{table-parameters} in Appendix~A. Then, the criteria of neutron tagging is determined by likelihood method based on these variables. The signal efficiency ($\varepsilon_{\rm sig,n}$) was estimated to be $\sim$17.7\%, while the probability that accidental background would be misidentified as a neutron ($\varepsilon_{\rm mis}$) was $\sim$1\%.

The algorithm was then updated to reach $\varepsilon_{\rm sig,n}\sim$20\% with the same background probability by using a machine learning model trained on an MC sample of 2.2~MeV $\gamma$ emission with a neutron capture time of $200~\rm \mu s$ \cite{SK2016spallation}.

In this work, we trained a new model on a MC sample of neutron emission from the solar $\bar{\nu}_e$. The vertex of neutron emission was distributed uniformly within the SK volume, and the neutron recoil and capture were taken into account in the MC sample. The characteristic variables, event selection, machine learning method, and performance evaluation are detailed in Appendix~A.

The previous search \cite{SK2015} suggests the dominant background is accidental coincidence with PMT dark noise. In order to suppress that background by a factor of 100, i.e. $\varepsilon_{\rm mis}\sim0.01\%$, we apply a tight cut with $\varepsilon_{\rm sig,n}=12.6\%$ (10.8\%) for a 535~$\rm \mu s$ (385~$\rm \mu s$) of the delayed-coincidence time window.

To evaluate the uncertainty on the absolute neutron tagging efficiency, we took neutron capture data in 2009 and 2016 using an AmBe calibration source \cite{SKAmBe2009} embedded in the center of a 5~cm cube of bismuth germanate oxide (BGO) scintillator. The AmBe source which emits neutrons was deployed at three different positions, labeled A (at the center of the detector), B (close to the barrel), and C (close to the top). The prompt signal of 700--1050 photoelectrons from scintillation light produced in the BGO, which corresponds to the 4.43~MeV $\gamma$ peak, is used to select an event sample with neutron emission from the source \cite{ATMPDntag}. This neutron is typically captured in hydrogen around the source point, and then 2.2~MeV $\gamma$ is emitted. The distribution of capture time $\Delta T$, which is the time difference between the prompt and delayed signal, is fitted with the shape $A_0 \exp( -\Delta T / \tau ) + A_1$, where $A_0$ is the amplitude of neutron emission candidates, $\tau$ is the capture time constant, and $A_1$ is an accidental background term. The absolute efficiency uncertainty is estimated as $(\varepsilon_{\rm sig,n}-\varepsilon_{\rm n})/\varepsilon_{\rm sig,n}$, where $\varepsilon_{\rm n}$ is the tagging efficiency normalized to the time window of 535~$\rm \mu s$ from the calibration condition, e.g. the uncertainty is estimated to be 10\% in a case of the sample of the source A (center) in 2016. Results for the three locations A, B and C are summarized in Table \ref{table-ambe-result}.

The maximum inconsistency between the measured absolute efficiency and simulation, 19\% relatively, dominates in the systematic uncertainty of neutron tagging efficiency, hence the systematic error is estimated to be that factor.

\begin{table*}[htb]
\footnotesize
\centering
\caption{Analysis results for calibration using an AmBe source at positions A, B and C in 2009 and 2016.}
\begin{tabular}{ llll cc } \hline
Source& $x$& $y$& $z$ & $\varepsilon_{\rm n}$ & ${(\varepsilon_{\rm sig,n}-\varepsilon_{\rm n})/\varepsilon_{\rm sig,n}}$ \\
\hline
(2009)\\
A& 35.3~cm &  $-$70.7~cm & 0~cm    & $10.8\pm0.2\%$ & 0.14 \\
B& 35.3~cm &  1210.9~cm& 0~cm    & $10.3\pm0.2\%$ & 0.19 \\
C& 35.3~cm &  $-$70.7~cm & 1500~cm & $11.2\pm0.2\%$ & 0.11 \\
\hline
(2016)\\
A& 35.3~cm & $-$70.7~cm &  0~cm    & $11.3\pm0.2\%$ &  0.10 \\
B& 35.3~cm & 1210.9~cm&  0~cm    & $11.0\pm0.2\%$ &  0.13 \\
C& 35.3~cm & $-$70.7~cm &  1500~cm & $11.1\pm0.2\%$ &  0.12 \\
\hline
\end{tabular}
\label{table-ambe-result}
\end{table*}

\subsection{Selected IBD events}
\label{sec.ev.ibd}

Table ~\ref{tab:summary.sparation} summarizes the cut criteria for each energy region and the number of surviving events in this analysis.

Since the signal's positron kinetic energy can reach up to ${\sim15~\rm MeV}$ when including detector resolution, we search in the region of 7.5--15.5~MeV in this analysis, as shown in Fig.~\ref{fig:data.profile.ene}. Finally 78 IBD candidates are obtained after the first, second, third, and fourth reduction.

\begin{figure}[htb!]
\centering
\includegraphics[width=3in]{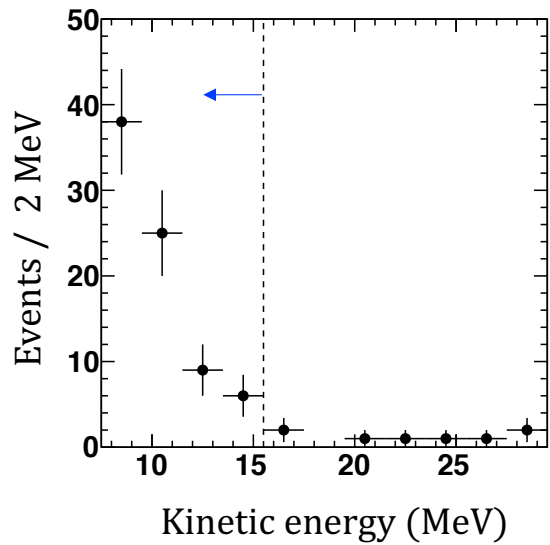}
\caption{Positron kinetic energy distribution. The searched region is 7.5--15.5~MeV to the left of the dashed line.}
\label{fig:data.profile.ene}
\end{figure}
    
The reconstructed vertex point distribution in the fiducial volume is shown in Fig.~\ref{fig:data.profile.vertex}. We have found no significant spatial cluster.

\begin{figure}[htb!]
\centering
\includegraphics[width=3in]{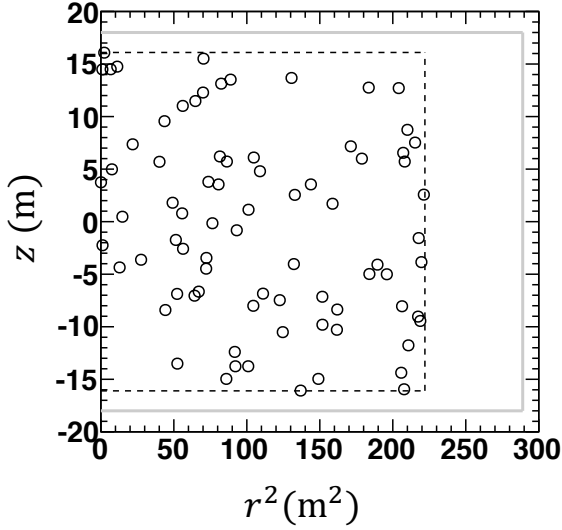}
\caption{Reconstructed vertex profile of prompt signal events. The dashed line indicates the fiducial cut region. The solid thick line is the boundary between ID and OD regions.}
\label{fig:data.profile.vertex}
\end{figure}
    
The time difference between a real neutron capture and the corresponding prompt signal ($\Delta T$) is fit by a function of $A_0 \exp(-\Delta T/\tau) + A_1$ as shown in Fig.~\ref{fig:data.profile.deltaT}, where it is fixed to a time constant of $\tau=204.8~{\rm \mu s}$ in assumption of signal of neutron capture in proton. The fitted parameters are $A_0 = 3.67 \pm 1.75$ and $A_1 = 2.27  \pm 0.64$ with $\chi^2/dof = 23.5/20$.

\begin{figure}[htb!]
\centering
\includegraphics[width=3in]{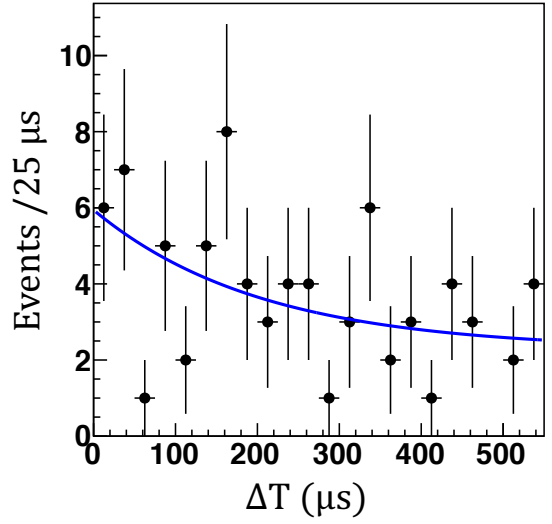}
\caption{Capture time $\Delta T$ distribution. The blue line is a fitting function of power law with time component of $204.8~\mu \rm s$.}
\label{fig:data.profile.deltaT}
\end{figure}

The event rate above the energy-threshold of 9.5~MeV (October 2008 to September 2011) or 7.5~MeV (September 2011 to May 2018) is shown as a function of time in Fig.~\ref{fig:data.profile.rate}. From September 2011 onward, the trigger threshold was lowered, then the event rate is shifted up due to the lower energy threshold; it is stable within each condition's statistical uncertainty.
    
\begin{figure}[htb!]
\centering
\includegraphics[width=3in]{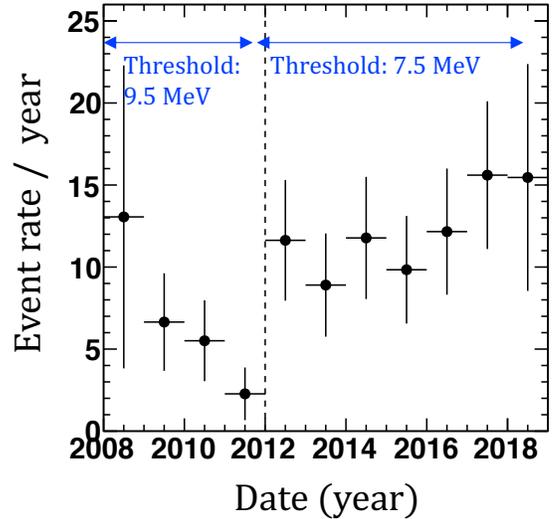}
\caption{Event rate as a function of time. The energy threshold was lowered from 9.5~MeV to 7.5~MeV in September 2011.}
\label{fig:data.profile.rate}
\end{figure}

\begin{table}[htb]
\centering
\footnotesize
\caption{The number of surviving events between 7.5--15.5~MeV ($N$) and signal efficiency for the 7.5--9.5, 9.5--11.5, and 11.5--19.5~MeV regions.
The bottom line indicates the total number of the survived events and all efficiency applying for the signal.
}
\label{tab:summary.sparation}
\begin{tabular}{l r rrr} 
\hline
Cut criteria & $N$ & \multicolumn{3}{c}{Signal Efficiency/ E(MeV)}\\
&& 7.5--9.5 & 9.5--11.5 & 11.5--19.5\\
\hline
First Reduction     & 1,404,568& 77.9\% & 80.1\% & 80.6\%\\
Spallation cut      & 213,576& 53.6\% & 55.2\% & 75.3\%\\
$effwall$ cut & 176,646& 91.6\% & 91.0\% & 90.1\%\\
Cherenkov angle cut & 88,778& 77.9\% & 83.2\% & 99.4\%\\
{\it pilike} cut    & 88,033& 98.8\% & 99.2\% & 99.4\%\\
Charge/Hit cut      & 87,372& 99.5\% & 99.7\% & 99.8\%\\
Neutron tag         & 78& 12.6\% & 12.6\% & 12.6\%\\
\hline
Total               & 78 &3.7\%&4.1\%&6.0\%\\
\hline
\end{tabular}
\end{table}

\section{Background estimation}

The background for solar $\bar{\nu}_e$ IBD events consists of atmospheric neutrinos, $^9$Li events, reactor $\bar{\nu}_e$, and accidental coincidences.

\subsection{Atmospheric neutrinos}

In this work, the atmospheric neutrino background is grouped into two categories: NCQE-like and non-NCQE. The former produces nuclear de-excitation $\gamma$-rays whose final state energy is $\mathcal{O}(10)~\rm MeV$, hence it could be mimicked. The latter is mainly made up of decay electrons which are produced by $\mu$--$e$ decay via muon neutrino charged-currents and by $\pi$--$\mu$--$e$ decay via neutrino neutral-currents with pion, where the muon and pion emit no Cherenkov photon.

A simulated sample of atmospheric neutrino events corresponding to 500~years of livetime is produced and then normalized to the SK-IV livetime. It is then scaled with a factor of $\varepsilon_{\rm sig,\mu}$ to account for the second reduction. It is processed with the third and fourth reduction and a kinetic energy threshold of 9.5~MeV or 7.5~MeV is applied, depending on the date.

The resulting NCQE-like sample consists of 7.8 events in the energy range of 7.5--15.5~MeV. The systematic error is evaluated to be ${+67.7\%}/{-65.6\%}$ by separately considering $\nu$ and $\bar{\nu}$ for the kinetic energy regions of 7.5--9.5, 9.5--11.5, and 11.5--15.5~MeV, and taking into account the cross section uncertainty reported by T2K~\cite{T2K2019}, the error of reduction cut efficiency, neutron tagging uncertainty, and neutron emission multiplicity.

The estimated number of background events in the non-NCQE simulation sample is evaluated using data in the higher kinetic energy region, 29.5--79.5~MeV. The dominant background consists of events with non-NCQE interactions by atmospheric neutrinos, which mainly accompany decay electrons. The surviving non-NCQE sample in the simulation is consistent with data, as shown in Fig.~\ref{fig.sideband.atm}. The simulation was a little lower than data, and it was considered the difference occurred from effects of several uncertainties such as model, flux, cross section, and cut efficiencies. In order to perform fine-tuning of number of non-NCQE sample in the 7.5--15.5~MeV region, the simulated spectrum was normalized to data in the 29.5--79.5~MeV region, as sideband analysis.
The correction factor of $1.17 \pm 0.15$ is determined and the corrected spectrum is shown in the red line on  Fig.~\ref{fig.sideband.atm}.

The corrected non-NCQE sample consists of 3.0 events in the 7.5--15.5~MeV region, with a systematic uncertainty of 12.8\% from the correction factor's error.

\begin{figure}[htb]
\includegraphics [width=3.5in]{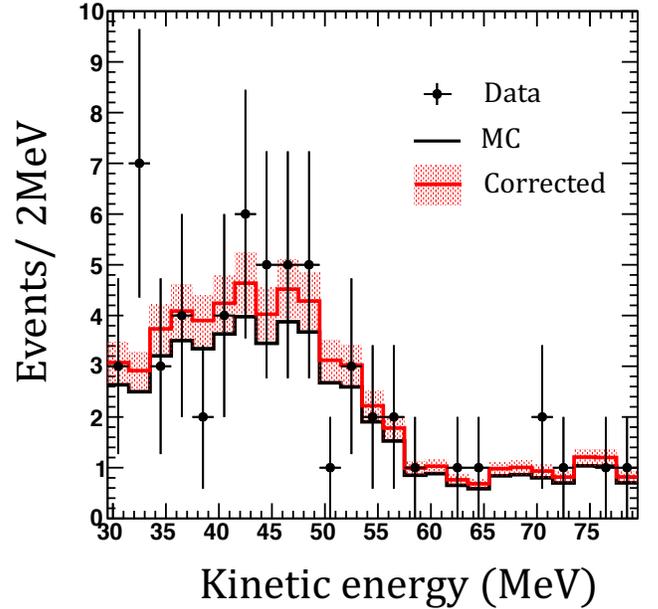}
\caption{Consistency between data and simulation of atmospheric neutrinos after the neutron-tagging reduction in the 29.5--79.5~MeV region. The black dots are data. The thick black line is the distribution of non-NCQE interaction: the charged current interaction (CC) and neutral current interaction (NC) with pion production. The red band is the corrected spectrum with fitting error, which is used to estimate the background of atmospheric neutrino events with non-NCQE interaction.}
\label{fig.sideband.atm}
\end{figure}

\subsection{{\rm$^9$Li} decay events}

A sample of ${\sim8.3\times10^6}$ $^9$Li decay events was simulated and then normalized to the predicted number of events by integrating $dN/dt$ of Eq.~\ref{eq.sim.li9} over the SK-IV livetime. It is then scaled with a factor of  $\varepsilon_{\rm Li9}$ to account for the second reduction. It is processed with the third and fourth reduction and a kinetic energy threshold of 9.5~MeV or 7.5~MeV is applied, depending on the date. The resulting $^9$Li decay sample consists of 40.0 events in the 7.5--15.5~MeV region.

The systematic error is evaluated to be 30\% by calculating the quadratic sum of the main factors: error of $Y_{\rm ^{9}Li}$, error of $Br$, reduction efficiency error, and the neutron tagging uncertainty.

\subsection{ Reactor $\bar{\nu}_e$ }
\label{sec.bg.reactor}

The $\bar{\nu}_e$ flux from nuclear reactors at the SK detector location has been estimated using the reactor database of the International Atomic Energy Agency \cite{IAEA_web}. The total reactor $\bar{\nu}_e$ flux for 10 years in SK is calculated to be $8.05\times10^{13} \rm cm^{-2}$.

Based on MC simulation, the kinetic energy spectrum of positrons from reactor $\bar{\nu}_e$ is derived via the re-weighting \cite{PRL.reweight} to the signal sample of $\bar{\nu}_e$ initial energy.

The simulated sample of $\bar{\nu}_e$ events is then normalized to the event number predicted by using the reactor total flux and the IBD cross section. Then, it is scaled with a factor of $\varepsilon_{\rm sig,\mu}$ to account for the second reduction. It is processed with the third and fourth reduction and a kinetic energy threshold of 9.5~MeV or 7.5~MeV is applied, depending on the date. The resulting reactor $\bar{\nu}_e$ sample consists of 1.2 events in the 7.5--15.5~MeV region. We adopt a 100\% systematic uncertainty as a conservative estimation, since the reactor database provides no uncertainties.

\subsection{Accidental coincidences}

The number of accidental coincidences between an electron or $\gamma$ and a dark noise fluctuation can be estimated by considering the time distribution of tagged neutrons in the sample after all event reduction processes. While the time difference $\Delta T$ of real neutron captures follows an exponential law, the time distribution for accidental coincidences is expected to be flat.

Furthermore, the number of accidental coincidences ($N_{\rm Accid}$) as a function of $\varepsilon_{\rm sig,n}$ is estimated by the same method. We empirically found that $N_{\rm Accid}$ tends to be a power law of $\varepsilon_{\rm sig,n}$ as shown in Fig.~\ref{relation-accidental-ntag}. The fit is used to determine the value of $N_{\rm Accid}$ for $\varepsilon_{\rm sig,n}=12.6\%$ as 41.9 events. The uncertainty of predicted number of accidental coincidence events is evaluated to be 27.7\%, arising from the error in the fit to the $\Delta T$ distribution.

\begin{figure}[htb]
\includegraphics[width=3.5in]{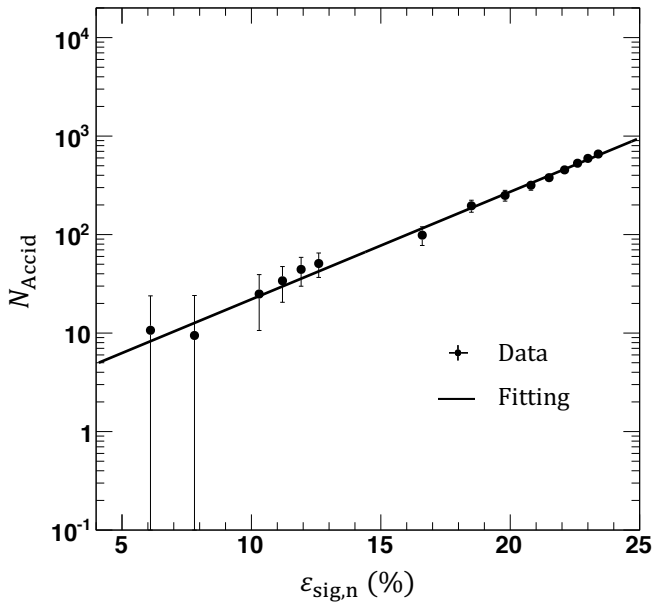}
\caption{Relation between signal tagging efficiency ($\varepsilon_{\rm sig,n}$) and the number of accidental coincidence events ($N_{\rm Accid}$). The fit is used to determine the value of $N_{\rm Accid}$.}
\label{relation-accidental-ntag}
\end{figure}

\subsection{Summary}

The predicted numbers of background events are summarized in Table~\ref{BG-estimation-summary}. For atmospheric neutrinos, $^9$Li decay events, reactor $\bar{\nu}_e$, and accidental coincidences, the errors of predicted number of events indicate systematic uncertainties to search for solar $\bar{\nu}_e$.

    \begin{table}[htb]
    \center
    \caption{
    Summary of the predicted numbers of background events in the kinetic energy region of 7.5--15.5~MeV for the whole livetime, 2970.1 live days.
    The time dependence of the energy thresholds is taken into account.
    }
    \begin{tabular}{ l r r r } \hline
    Source & Number of predicted events \\ \hline
    Atmospheric neutrinos\\
    ~~ NCQE-like interactions & $7.8^{+5.4}_{-5.2}$\\
    ~~ non-NCQE interactions & $3.0 \pm 0.4$ \\
    $^9$Li decay events & $40.0 \pm 12.0$ \\
    Reactor $\bar{\nu}_e$ & $1.2 \pm 1.2$ \\
    Accidental coincidences & $41.9 \pm 11.6$ \\
    \hline
    Total & $95.0 \pm17.6$ \\
    \hline
    \end{tabular}
    \label{BG-estimation-summary}
    \end{table}
\section{Analysis and results}

To search for solar $\bar{\nu}_e$ events with a positron kinetic energy in the 7.5--15.5~MeV region---equivalent to 9.3--17.3~MeV neutrino energy---the energy spectrum of IBD candidates is compared with the background estimation. We expected $95.0 \pm17.6$ events from the backgrounds and observed 78 events in data. The selected sample is consistent with background predictions and no significant signal was found. 

Figure~\ref{fig.result.spectrum} of cyan dashed line shows a predicted spectrum of kinetic energy for solar-$\bar{\nu}_e$ events in an assumption of $10^{-4}$ of the neutrino-to-antineutrino conversion provability. In this analysis, the number of solar-$\bar{\nu}_e$ events is derived after the fitting with the signal and background spectra. The observed numbers of events for the four energy bins are compared to the best-fit signal and background predictions. The amplitude of the signal is a free parameter. The signal and backgrounds have a known spectral shape which is included in the fit. Therefore, the upper limit of the conversion probability is evaluated in this study. In addition, it is enough to determine the limit based on maximum likelihood with $\Delta\chi^2$ test because of the simple fitting in this case.

\begin{figure}[htb]
\includegraphics[width=3.3in]{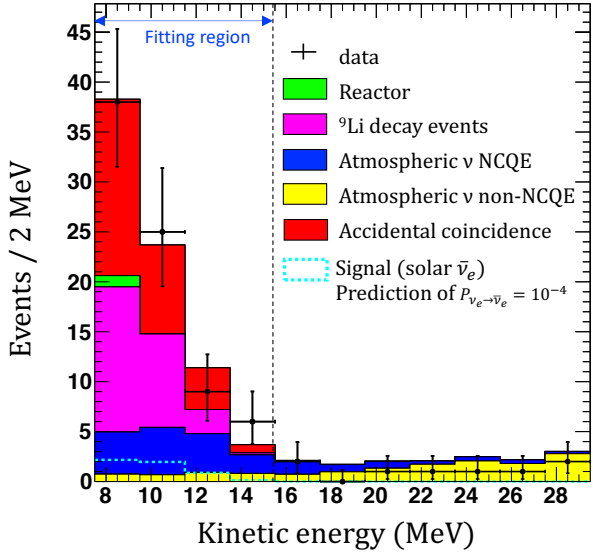}
\caption{Fit result in the kinetic energy range of 7.5--15.5~MeV. The black dots are data. The green, magenta, blue, yellow, and red histograms show best-fit predictions for reactor antineutrino events, $^9$Li decay events, atmospheric neutrino's NCQE interactions and non-NCQE interactions, and accidental coincidences, respectively. The cyan dashed line is solar antineutrino signal events in an assumption of $10^{-4}$ of a neutrino-to-antineutrino conversion probability.}
\label{fig.result.spectrum}
\end{figure}

In order to evaluate the conversion provability $P_{\nu_e\to\bar{\nu}_e}$, the $\chi^2$ is defined as,
\begin{eqnarray}
    \chi^2 = 2
    \sum_{j}
    \left(
    \mu_j - n_j + n_j \ln \frac{n_j}{\mu_j} 
    \right)
    + \sum_{k}
    {\left(\frac{\alpha_k}{\sigma_k}\right)^2},
    \label{eq-chi2}
\end{eqnarray}
where $\mu_j$ and $n_j$ are the predicted and observed number of events in the $j^{\rm th}$ energy bin, respectively. The second term in Eq.~\ref{eq-chi2} is a pull term for the six free parameters ($\alpha_k$) which model the uncertainties on the different event and the fractional errors of predicted number of events ($\sigma_k$), where the index $k$ corresponds to the category: (1) NCQE-like and (2) non-NCQE interactions of atmospheric neutrinos, (3) $^9$Li decay events, (4) reactor $\bar{\nu}_e$, (5) accidental coincidence, and (6) solar $\bar{\nu}_e$ signal. The predicted $\mu_j$ is a function of the simulated signal and background estimation given as,
        \begin{eqnarray}
    \mu_j &=& \sum^{5}_{k=1}
    (1+\alpha_k) (1+\omega_{k, j} ) N_{k, j}  \nonumber \\
    &&+ P_{\nu_e\to\bar{\nu}_e} (1+\alpha_6) (1+\omega_{6, j}) N_{6, j},
    \end{eqnarray}
where $N_{k, j}$ is the predicted number of background and signal events. The $\sigma_k$ for background is a prediction error as listed in Table \ref{BG-estimation-summary}. In particular, $N_{6,j}$ indicates the predicted number of the solar $\bar{\nu}_e$ events in an assumption of $P_{{\nu_e}\to\bar{\nu}_e}=1$. The $\sigma_6$ is estimated to be 20\% for the number of solar $\bar{\nu}_e$ events by calculating the quadratic sum of the reduction efficiency error, the neutron tagging uncertainty, and a time depending uncertainty of the efficiency. Therefore, the correlation between $P_{{\nu_e}\to\bar{\nu}_e}$ and $\alpha_6$ has been taken into account. The $\omega_{k, j}$ parameterize spectral shape distortions and can go up to 10\% for atmospheric neutrino NCQE interactions and accidental coincidences. Other spectral shape distortions can be up to 1\% because the 2-MeV~bin is larger than the energy resolution of the SK detector and spectral shapes for these backgrounds are well known. Assuming constant $P_{{\nu_e}\to\bar{\nu}_e}$ for any energy, the spectral shape of signal is set to a same of $^8$B solar $\nu_e$. The pull term of $\omega_{k,j}$ is a negligibly small effect in Eq.~\ref{eq-chi2}, it is omitted, because the parameter make distortion spectrum for signal or background but the total number of events is not changed. 

Thus, Eq.~\ref{eq-chi2} indicates the pull term contributions and that the background spectral shape predictions are providing constraints in the $\chi^2$ fit.
Since we have four bins, six parameters for systematic uncertainties in the pull term to the fit, and seven free parameters ($P_{\nu_e\to\bar{\nu}_e}$ and $\alpha_k$), the number of degrees of freedom ($dof$) is found to be equal to three.

    \begin{table}[htb]
    \center
    \caption{Best-fit $\alpha_k$ values}
    \begin{tabular}{ llr } \hline
    $k$ & Source & $\alpha_k$ \\
    \hline
    1 & NCQE-like interaction & $0.38 \sigma_1$\\
    2 & non-NCQE interaction& $\sigma_2$\\
    3 & $^{9}$Li decay events & $ 0.99 \sigma_3$\\
    4 & Reactor $\bar{\nu}_e$ & $-0.01 \sigma_4$\\
    5 & Accidental coincidence &$-0.61 \sigma_5$\\
    6 & Solar $\bar{\nu}_e$ & $0$\\
    \hline
    \end{tabular}
    \label{tab.alpha}
    \end{table}
    
\begin{figure}[htb]
\includegraphics[width=3.3in]{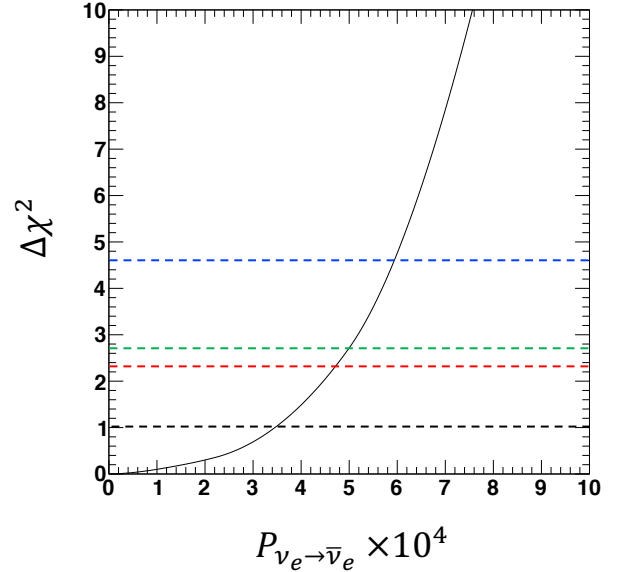}
\caption{
Relation between $\Delta\chi^2$ and conversion probability of neutrinos to antineutrinos. The upper limits on $P_{\nu_e\to\bar{\nu}_e}$ for $\Delta \chi^2=1.0,~2.3,~2.7,~\rm and~4.6$ are $3.5\times10^{-4}$, $4.7\times10^{-4}$, $5.0\times10^{-4}$, and $6.0\times10^{-4}$, respectively.}
\label{fig.result.chi2}
\end{figure}

The best-fit conversion probability is $P_{\nu_e\to\bar{\nu}_e}^{(\rm best)}=0$ and corresponds to the value of $\alpha_k$ listed to Table \ref{tab.alpha}. The best-fit $\chi^2/{dof}$ is $1.9/3$ (shown in Fig.~\ref{fig.result.spectrum}) under the null hypothesis. The p-value is 0.6 and no significant signal above the backgrounds is found. In the $\Delta \chi^2$ calculation with an increased value of $P_{\nu_e\to\bar{\nu}_e}$, the $\alpha_k$ parameters are also varied within uncertainties associated with backgrounds, as shown in Fig.~\ref{fig.result.chi2}. Then, requiring $\Delta \chi^2<2.3$, the upper limit is determined to be $P_{\nu_e\to\bar{\nu}_e}<4.7\times10^{-4}$ at 90\%~C.L., which corresponding to 36 events of solar-$\bar{\nu}_e$ signal. The $^{8}$B neutrino flux from the Sun above 9.3~MeV of neutrino energy is calculated as $9.96 \times 10^5~{\rm cm^{-2}~s^{-1}}$; therefore the partial flux upper limit of antineutrino from the Sun is determined to be $4.7\times10^2~{\rm cm^{-2}~s^{-1}}$ at 90\%~C.L. The neutrino magnetic moment derived from the $\nu_e\to\bar{\nu}_e$ probability in the spin-flavour precession model is calculated as $\mu \lesssim 1.7\times10^{-9}\mu_B (10~{\rm kG}/B_T)$ at 90\%~C.L., i.e. $\mu \lesssim 3\times10^{-8}\mu_B$ and $\mu \lesssim 2\times10^{-12}\mu_B$ at 90\%~C.L. in the assumption of ${B_T \sim600~\rm G}$ \cite{Kitchatinov2008} and ${\sim7~\rm MG}$ \cite{Friedland2003}, respectively.
\section{Sensitivity estimate for SK-Gd}

The sensitivity of SK-Gd can be estimated using the SK-IV result in the neutrino energy range of 9.3--17.3~MeV. We expect the sensitivity of the solar $\bar{\nu}_e$ search to significantly improve in SK-Gd, since this upgrade considerably improves the neutron identification efficiency.

In this work, we find that the main backgrounds in SK are $^9$Li decay events, atmospheric neutrino NCQE interaction, and accidental coincidences. The estimated SK-Gd sensitivity is summarized in Table~\ref{tab.sum.SKGD}.

    \begin{table*}[htb]
    \center
    \caption{Expected sensitivity of SK-Gd for $P_{\nu_e\to\bar{\nu}_e}$ at 90\%~C.L. in an assumption of 10 years observation.}
    \begin{tabular}{ l c c  } \hline
    & \multicolumn{2}{c}{$P_{\nu_e\to\bar{\nu}_e}$}\\
    & 0.02\% Gd loading & 0.2\% Gd loading\\
    \hline
    Improved signal efficiency
    & $10.5\times10^{-5}$& $5.9\times10^{-5}$\\
    + $^9$Li rejection (to 20\%)
    & $8.3\times10^{-5}$& $4.6\times10^{-5}$\\
    + Accidental coincidence rejection (to 5\%)
    & $5.6\times10^{-5}$& $3.1\times10^{-5}$\\
    + NCQE uncertainty decreasing (to 30-80\%)
    & $(4.3-5.2)\times10^{-5}$& $(2.4-2.9)\times10^{-5}$\\
    \hline
    \end{tabular}
    \label{tab.sum.SKGD}
    \end{table*}

First, the signal efficiency is evaluated to be 50\% and 90\% for 0.02\% and 0.2\% Gd sulfate loading, respectively \cite{SKGd2016}. After 10 years of observation, the sensitivity to $P_{\nu_e\to\bar{\nu}_e}$ at 90\%~C.L. is expected to be $10.5\times10^{-5}$ ($5.9\times10^{-5}$) when accounting only for the efficiency improvement to 50\% (90\%), and assuming that the uncertainty of the neutron tagging efficiency is reduced from 19\% to 5\%.

Additionally, due to the improved detection efficiency for neutron tagging, SK-Gd would measure the yield of $^9$Li decay events more precisely, i.e. the $^9$Li decay events can be observed more than  Ref.~\cite{SK2016spallation} and expected to investigate the spallation mechanism precisely. After that, when a better separation cut condition is developed, hence the $^9$Li background can be reduced for the search in SK-Gd. In this estimate, it is assumed to improve the total uncertainty for $^{9}$Li from 30\% to 5\% and the spallation efficiency $\varepsilon_{\rm Li9}$ to 20\% of its current value. These resulting sensitivities are predicted to be $8.3\times10^{-5}$ (0.02\% Gd loading) and $4.6\times10^{-5}$ (0.2\% Gd loading).

The accidental coincidences consist of (1) fake tagging and (2) real neutron capture of unrelated neutrons. In SK-Gd, the fake tagging will be considerably reduced since the energy of $\gamma$ after neutron capture increases from 2.2~MeV in hydrogen to 8~MeV in Gd. On the other hand, accidental coincidences of unrelated, real neutrons cannot be identified by the tagging algorithm, even in SK-Gd. Accidental coincidences should be precisely evaluated during calibration with random triggering in SK, both with and without Gd, which will allow us to reduce the associated uncertainties. Assuming a 5\% uncertainty, we expect the resulting sensitivity to be $5.6\times10^{-5}$ (0.02\% Gd loading) and $3.1\times10^{-5}$ (0.2\% Gd loading), where the estimates include the contribution of improved $^{9}$Li rejection.

As the remaining factor for further improvement of the sensitivity, the uncertainty of the spectrum of atmospheric neutrino NCQE interactions is expected to be reduced by a future artificial neutrino beam experiment. In particular, the current uncertainty is estimated to be $\sim$100\%, considering both the MC prediction error and the shape uncertainty of the spectrum in the low energy region. If the uncertainty can be decreased to 30--80\% of the present one, the resulting sensitivities are calculated to be (4.3--5.2)$\times10^{-5}$ (0.02\% Gd loading) and (2.4--2.9)$\times10^{-5}$ (0.2\% Gd loading), where the estimates include the contribution of improved rejection of both $^{9}$Li and accidental coincidences. This estimate indicates an improvement of a factor of $\sim16$ from the present sensitivity for conversion probability, which would make it possible to improve upon the current best upper limit set by other experiments.

Finally, the sensitivity of SK-Gd could be improved by lowering the energy threshold. The trigger condition should be tuned in consideration of the allowed dark rate contamination after the gadolinium loading.
\section{Conclusion}
We searched in the SK detector for solar $\bar{\nu}_e$ due to $\nu_e\to\bar{\nu}_e$ conversion, using neutron tagging to identify IBD interactions in pure water. The selected sample is consistent with background predictions and no significant signal was found. An upper limit on the $\nu_e\to\bar{\nu}_e$ conversion probability of $4.7\times10^{-4}$ is hence derived at 90\%~C.L. This limit is a factor of 17 more stringent than the SK-I sensitivity and is consistent with the sensitivity estimated at a previous search in SK-IV \cite{SK2015}. This limit corresponds to the neutrino magnetic moment of $\lesssim 1.7\times10^{-9}\mu_B (10~{\rm kG}/B_T)$ at 90\%~C.L. predicted in the spin-flavour precession model. This SK-IV analysis thus derived the best limit of sensitivity to solar $\bar{\nu}_e$s at SK and has allowed us to assess the 16 times improvement from the present sensitivity expected for future searches in SK-Gd.
\section*{Acknowledgment}

We gratefully acknowledge the cooperation of the Kamioka Mining and Smelting Company.
The Super‐Kamiokande experiment has been built and operated from funding by the Japanese Ministry of Education, Culture, Sports, Science and Technology, the U.S.
Department of Energy, and the U.S. National Science Foundation.
Some of us have been supported 
by funds from 
the National Research Foundation of Korea NRF‐2009‐0083526
(KNRC) funded by the Ministry of Science, ICT,
and Future Planning and the Ministry of
Education (2018R1D1A3B07050696, 2018R1D1A1B07049158), 
the Japan Society for the Promotion of Science,
the National Natural Science Foundation of China under Grants No. 11620101004,
the Spanish Ministry of Science, Universities and Innovation (grant PGC2018-099388-B-I00), 
the Natural Sciences and Engineering Research Council (NSERC) of Canada, 
the Scinet and Westgrid consortia of Compute Canada, 
the National Science Centre, Poland (2015/18/E/ST2/00758),
the Science and Technology Facilities Council (STFC) and GridPPP, UK, 
the European Union's  Horizon 2020 Research and Innovation Programme under the Marie Sklodowska-Curie grant
agreement no.754496,
H2020-MSCA-RISE-2018 JENNIFER2 grant agreement no.822070, 
and H2020-MSCA-RISE-2019 SK2HK grant agreement no. 872549.
This analysis was supported also by KAKENHI Grant-in-Aid for Scientific Research (C) No. 20K03998.
\section*{Appendix A: Neutron tagging
\label{sec.tmva}
}

In what follows, we detail the structure of the neutron tagging algorithm used for this analysis and evaluate its performance.

    \begin{figure*}[hbt]
    \centering
    \includegraphics[width=8in]{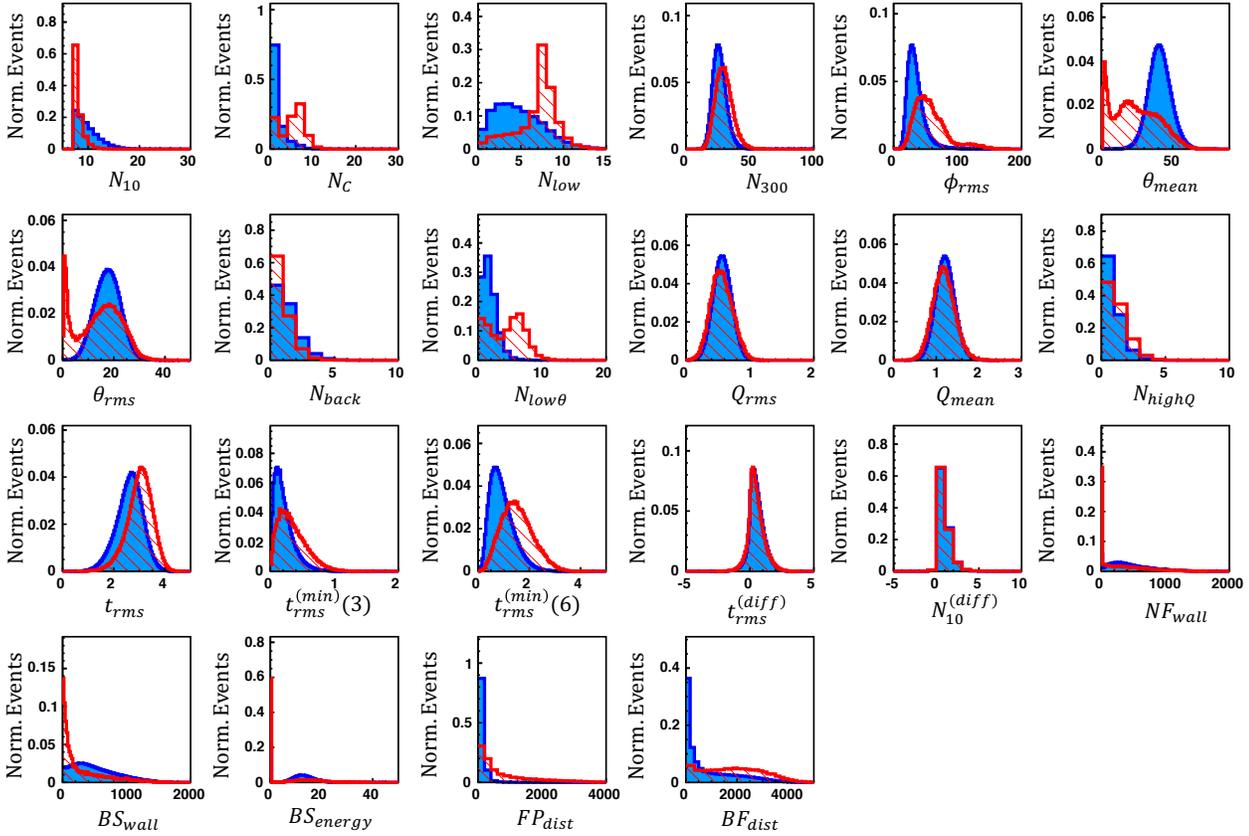}
    \caption{Variable distributions used for neutron tagging. The blue filled and red hatched histogram correspond to the 2.2~MeV $\gamma$ signal and background PMT dark noise, respectively. Both histograms are normalized to 1.}
    \label{ntag-param}
    \end{figure*}

\subsection*{A.1. Characteristic variables}
 
A neutron from IBD can be identified by tagging the 2.2~MeV $\gamma$ emission resulting from its capture on hydrogen \cite{SK2015}. In this analysis, the simulation signature which is merged with the T2K dummy data allows to accurately model the contributions of the different backgrounds. Since the $\gamma$ signal is typically hidden under the PMT dark noise, we use the the number of hits together with their timing, pattern, and charge to discriminate between the signal of the 2.2~MeV $\gamma$ and background PMT dark noise, as shown in Fig.~\ref{ntag-param}.

The definitions of some variables are presented in Ref.~\cite{SK2015, SK2016spallation, SK-D-thesis-2014, SK-D-thesis-2018} and summarized in Table \ref{table-parameters}. Basically, we take the hit clusters within a 10~ns window in AFT data to calculate quantities such as the hit number, timing deviation, and angle. To identify neutron capture events, we apply two approaches for vertex reconstruction using the selected candidates in a 10~ns window. One is based on minimal root-mean-square (RMS) for mean time-of-flight (TOF) between the vertex and the hit PMT position for the candidates in the 10~ns window \cite{SK-D-thesis-2014} and the other is based on the minimal timing residual defined as the difference between a PMT’s observed hit time and the expected hit time based on the time-of-flight of the Cherenkov photon~\cite{SK2008-solar}. The first method is used to derive the vertex ($\vec{x'}$) and the minimal RMS of time ($t'_{\rm rms}$), while the primary vertex is labeled $\vec{x}$. In addition, the latter method is used to derive the vertex ($\vec{x'_b}$) and the charge ($Q'_{b}$).

We then use a machine learning algorithm (explained later) to identify neutron capture events efficiently using the correlation of these variables.

    \begin{table}[htb]
    \footnotesize
    \centering
    \caption{Overview of parameters used for neutron tagging.}
    \begin{tabular}{ l l  } \hline
    Parameter & Meaning \\
    \hline
    $N_{10}$ & Number of PMTs hit in a 10 ns window.\\  
    $N_{\rm C}$ & Number of clusters among $N_{10}$ candidates.\\   
    $N_{\rm low}$ & Hit number on low probability \cite{SK2015}.\\
    $N_{300}$   & Hit number in 300 ns width at timing center of $N_{10}$. \\
    $\phi_{\rm rms}$& RMS of azimuthal angle of the vectors.\\
    $\theta_{\rm mean}$ & Mean of opening angle between vectors of each PMT  \\& and  sum of all. \\   
    $\theta_{\rm rms}$  & RMS of the opening angle.\\   
    $N_{\rm back}$      & Number of hits with $\theta>90^{\circ}$. \\   
    $N_{\rm low\theta}$ &  Number of hits with $\theta<20^{\circ}$.\\
    $Q_{\rm rms}$       & RMS of charge.\\
    $Q_{\rm mean}$      & Mean of charge.\\
    $N_{\rm highQ}$     & Number of hits with high charge, $Q>3~\rm p.e.$\\   
    $t_{\rm rms}$ & RMS of hit time within 10-ns hit candidates.\\   
    $t_{\rm rms}^{\rm (min)}(3)$ & Minimum RMS of hit time with 3 hit PMTs. \\
    $t_{\rm rms}^{(\rm min)}(6)$ & Minimum RMS of hit time with 6 hit PMTs. \\
    $t_{\rm rms}^{(\rm diff)}$ & Difference between $t_{\rm rms}$ and $t'_{\rm rms}$. \\ 
    $N_{10}^{(\rm diff)}$ & Difference between $N_{10}$ and $N'_{10}$.\\
    ${NF}_{\rm wall}$ & Distance from wall to $\vec{x'}$.\\
    ${BS}_{\rm wall}$ & Distance from wall to $\vec{x'_b}$. \\
    ${BS}_{\rm energy}$ & Reconstructed energy based on $Q'_b$.\\   
    ${FP}_{\rm dist}$ & Distance between $\vec{x}$ and $\vec{ x'}$.\\
    ${BF}_{\rm dist}$ & Distance between $\vec{x'}$ and $\vec{x'_b}$.\\   
    \hline
    \end{tabular}
    \label{table-parameters}
    \end{table}
    
\subsection*{A.2. Pre-selection}
For each event associated with an SHE+AFT trigger pair, we look for neutrons in a $535~\rm \mu s$ ($385~\rm \mu s$) window. Since this window contains an extremely large number of timing hit clusters, we apply a pre-selection cut to suppress  huge background to speed up calculation. As a pre-selection cut, we consider 10-ns TOF-subtracted windows containing more than 7 hits --- $N_{10} > 7$. In addition, the criterion of $N_{300} - N_{10} > 8$ is required, where the PMT noise tends to be distributed randomly in time so the number of hits in a 300~ns window around the 10~ns window is a good index to confirm it. The pre-selection efficiency is estimated to be 80.4\% for simulated neutron-capture events while 65.6\% of the background dark noise events which are sampled from T2K dummy data are suppressed by this pre-selection procedure.

\subsection*{A.3. Machine learning} In order to select neutron-capture events from pre-selected candidates, we use a feed-forward Multi-Layer Perceptron (MLP) implemented in the TMVA library of ROOT \cite{ROOT-TMVA-2010}. In this analysis, the MLP was trained using $1.2\times10^6$ simulated neutron-capture events and $1.2\times10^7$ background triggers from the T2K dummy data. These were split randomly into a training sample of 75\% and a evaluation sample of 25\%.

\subsection*{A.4. Performance evaluation}

Using the MLP likelihood profiles of MC and background, the relation between the signal tagging efficiency ($\varepsilon_{\rm sig,n}$) and the accidental background probability ($\varepsilon_{\rm mis}$) are estimated as shown in Fig.~\ref{ntag-eff-curve}. $\varepsilon_{\rm mis}$ is defined as the probability that a sample containing only PMT dark noise is misidentified as a neutron capture event. The black dot marks the working point, which was selected based on the criteria described in Section~\ref{sec.event.4th}.

The SK electronics were updated to extend the delayed-coincidence time window from $\rm 385~\mu s$ to $\rm 535~\mu s$ in November 2008 and reduced the visible energy threshold of the prompt signal from 9.5~MeV to 7.5~MeV in September 2011. The neutron tagging efficiency was also affected by the DAQ upgrade. The signal efficiency in a 385 $\rm \mu s$ time window is estimated to be $0.86$ times as high as in a 535~$\rm \mu s$ time window.

The background dark noise event rate has a time dependence due to the PMT gain shift and water transparency fluctuation. In the simulation, T2K dummy data from 13$^{\rm th}$ March to 1$^{\rm st}$ November, 2009 was used. We confirmed that the neutron-tagging efficiency depends on time via the background, as a contribution from PMT gain shift almost, thus the shift of signal-tagging efficiency is estimated to be $+0.047\%/\rm year$ while the fake-tagging rate tends to increase by 2.5\% per year. This efficiency shift is treated as systematic uncertainty of $\sim4\%$ over the whole SK-IV livetime, but its value is negligible compared to the systematic error contributing to the absolute efficiency uncertainty, as explained in section \ref{sec.event.4th}.

    \begin{figure}[htb]
    \includegraphics[width=4 cm, bb= 0 150 200 520]{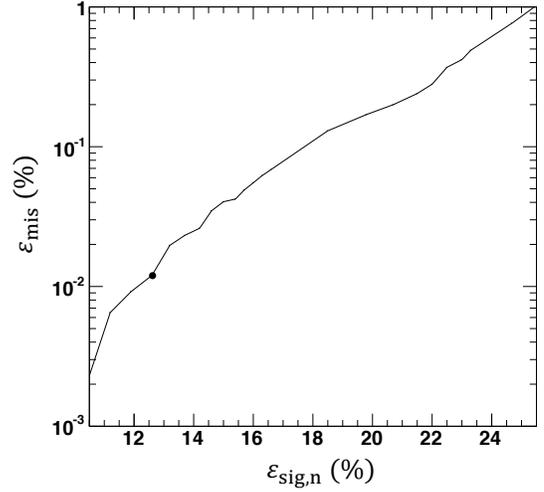}
    \caption{Relation between the signal efficiency ($\varepsilon_{\rm sig,n}$) and the probability that accidental background would be misidentified as a neutron ($\varepsilon_{\rm mis}$) for neutron tagging in a 535~$\rm \mu s$ time window. The black dot marks the working point.}
    \label{ntag-eff-curve}
    \end{figure}

\end{twocolumn}
\end{document}